\documentstyle[sprocl,epsfig]{article}

\def\Journal#1#2#3#4{{#1} {\bf #2}, #3 (#4)}

\def\NPB{{\em Nucl. Phys.} B}
\def\PLB{{\em Phys. Lett.}  B}
\def\PRL{\em Phys. Rev. Lett.}
\def\PRD{{\em Phys. Rev.} D}
\def\ZPC{{\em Z. Phys.} C}
\def\CPC{\em Comp. Phys. Comm.}

\def\be{\begin{equation}}
\def\ee{\end{equation}}
\def\bea{\begin{eqnarray}}
\def\eea{\end{eqnarray}}

\def\etjet{E_T^{jet}}
\def\etajet{\eta^{jet}}

\def\etaphi{\eta-\varphi}
 
\def\rs{R_{SEP}}
 
\def\etar{-1<\etajet<2}

\def\q2{Q^2}
\def\pb1{pb$^{-1}$}

\def\g2{GeV$^2$}
 
\def\rr1{R=1}
\def\r7{R=0.7}
\def\R71{R=0.7\ {\rm and}\ 1}

\begin{document}

\title{JET CROSS SECTIONS AND JET SHAPES IN PHOTOPRODUCTION AT HERA$^\dag$
\footnotetext{$^\dag$Talk given at the ``Meeting on Deep Inelastic 
Scattering'', Madrid, Spain, June 1997.}}

\author{C. GLASMAN, REPRESENTING THE H1 AND ZEUS COLLABORATIONS}

\address{DESY - F1, Notkestrasse 85,\\ 22603 Hamburg, Germany}

\maketitle\abstracts{
Studies on the structure of the photon are presented by means of the extraction
of a leading order effective parton distribution in the photon and measurements
of inclusive jet differential cross sections in photoproduction. Measurements
of the internal structure of jets have been performed and are also presented
as a function of the transverse energy and pseudorapidity of the jets.}

The main source of jets at HERA \footnote{HERA provides collisions between
positrons of energy $E_e=27.5$ GeV and protons of energy $E_p=820$ GeV.} comes
from collisions between protons and quasi-real photons ($\q2\approx 0$, where
$\q2$ is the virtuality of the photon) emitted by the electron beam
(photoproduction). At lowest order QCD, two hard scattering processes
contribute to jet production \cite{p:owens}: the resolved process, in which a
parton from the photon interacts with a parton from the proton, producing two
jets in the final state; and the direct process, in which the photon interacts
pointlike with a parton from the proton, also producing two jets in the final
state.

The study of high-$p_T$ jet photoproduction provides tests of QCD and allows
to probe the structure of the photon. In perturbative QCD the cross section
for jet production is given by

\begin{equation}
{d^4\sigma\over dy dx_{\gamma} dx_p d\cos\theta^*}\propto
{f_{\gamma /e}\over y}\sum_{ij} {f_{i/\gamma}\over x_{\gamma}}
{f_{j/p}\over x_p} \vert M_{ij}(\cos\theta^*)\vert^2
\label{eq:one}
\end{equation}

where $f_{\gamma /e}$ is the flux of photons from the electron approximated by
the Weizs\"{a}cker-Williams formula; $y=E_{\gamma}/E_e$ is the inelasticity
parameter; $f_{i/\gamma}\ (f_{j/p})$ are the parton densities in the photon
\footnote{The resolved and direct processes are included in $f_{i/\gamma}$.}
(proton), extracted from the data; $x_{\gamma}=E_i/E_{\gamma}\ (x_p=E_j/E_p)$
is the fractional momentum of the incoming parton from the photon (proton);
and $\vert M_{ij}(\cos\theta^*)\vert^2$ are the QCD matrix elements for the
parton-parton scattering.

There are two approaches to the study of the structure of the photon. One
approach is to extract directly from the data an effective parton distribution
in the photon. The second approach is to measure jet cross sections that can
be calculated theoretically which then provide a testing ground for
parametrisations of the photon structure function.

The method \cite{p:comb} to extract an effective parton distribution in the
photon is based on the use of the leading order (LO) matrix elements for the
subprocesses with gluon exchange, which give the dominant contribution to the
jet cross section in resolved processes for the kinematic regime studied.
Then, the quark and gluon densities are combined into an effective parton
distribution:

\begin{equation}
\sum [ f_{q/{\gamma}}(x_{\gamma},p_T^2)+
f_{\bar q/{\gamma}}(x_{\gamma},p_T^2) ] +
{9\over 4}f_{g/{\gamma}}(x_{\gamma},p_T^2)
\label{eq:two}
\end{equation}

\begin{figure}
\vspace{-1cm}
\centerline{\mbox{
\epsfig{figure=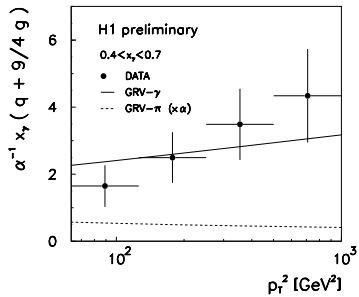,width=5cm}}}
\vspace{-0.5cm}
\caption{Leading order effective parton distribution in the photon.
\label{fig:one}}
\vspace{-0.5cm}
\end{figure}

A LO effective parton distribution in the photon was extracted \cite{p:h1} by
unfolding the double differential dijet cross section as a function of the
average transverse energy of the two jets with highest transverse energy and
of the fraction of the photon's energy participating in the production of the
two highest-transverse-energy jets ($x_{\gamma}$). The measurement was
performed using a fixed cone algorithm \cite{p:cone} in the pseudorapidity
\footnote{The pseudorapidity is defined as $\eta=-\ln(\tan\frac{\theta}{2})$,
where the polar angle $\theta$ is taken with respect to the proton beam
direction.} ($\eta$) $-$ azimuth ($\varphi$) plane with $\r7$ in the kinematic
range $0.2<y<0.83$ and $\q2<4$ \g2. Figure \ref{fig:one} shows the extracted
effective parton distribution as a function of the scale $p_T$ (the transverse
momentum of the parton). The data exhibit an increase with the scale $p_T$
which is compatible with the logarithmic increase predicted by QCD
\cite{p:witten} (i.e., the anomalous \footnote{The anomalous component of
the photon reflects the contribution of quarks from the pointlike coupling of
the photon to a quark-antiquark pair.} component of the photon structure
function). The predictions using the GRV-LO \cite{p:grvlo} parametrisations
of the parton densities in the photon and the pion are also shown in figure
\ref{fig:one}. The data disfavour a purely hadronic behaviour and are
compatible with the prediction which includes the anomalous component.

The second approach to the study of the photon structure function is to provide
measurements of differential jet cross sections at high jet transverse
energies, where the proton parton densities are constrained by other
measurements, and, therefore, are sensitive to the photon structure function.

\begin{figure}
\vspace{-1cm}
\centerline{\mbox{
\epsfig{figure=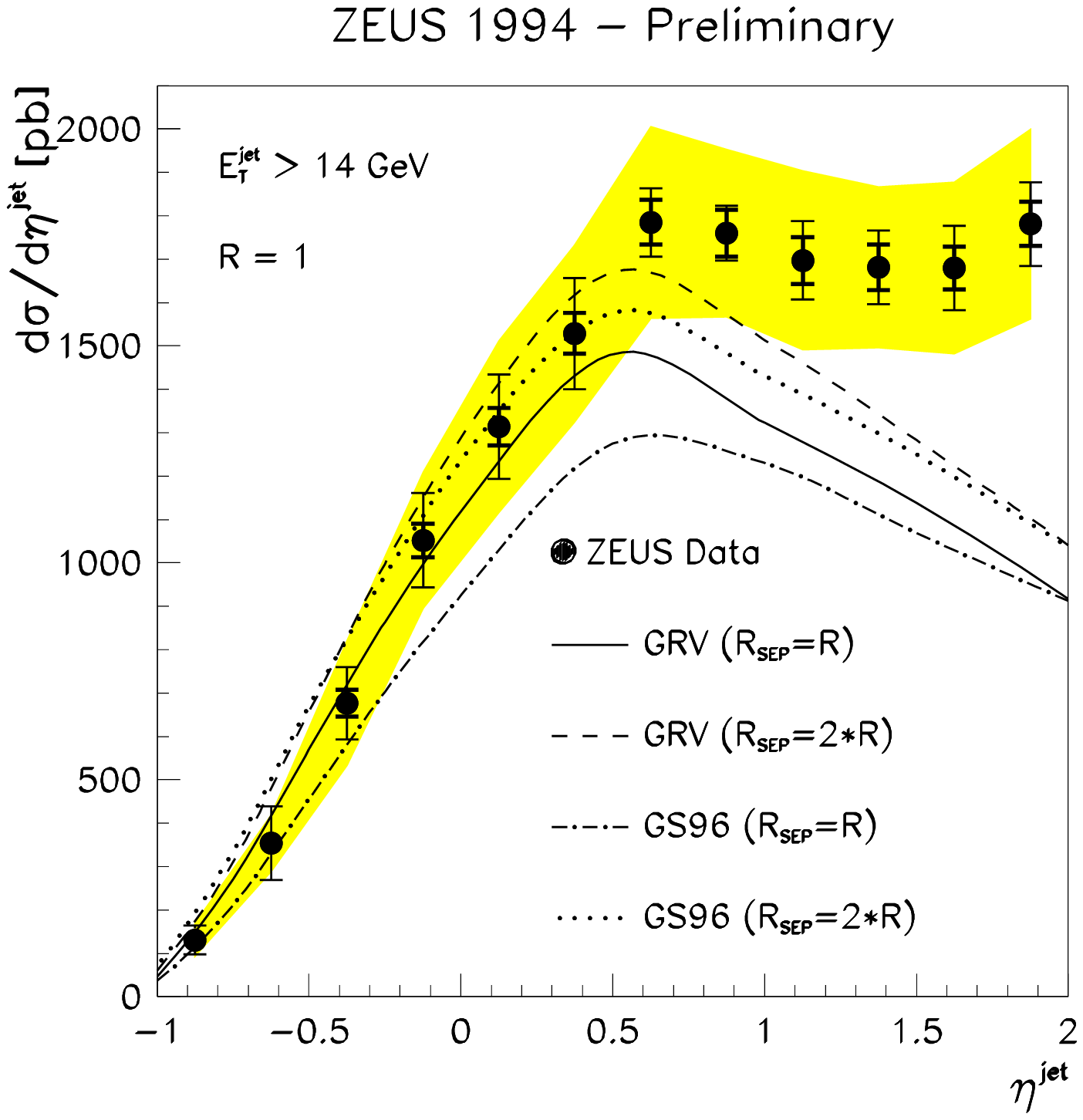,width=7.5cm}
\hspace{-2cm}
\epsfig{figure=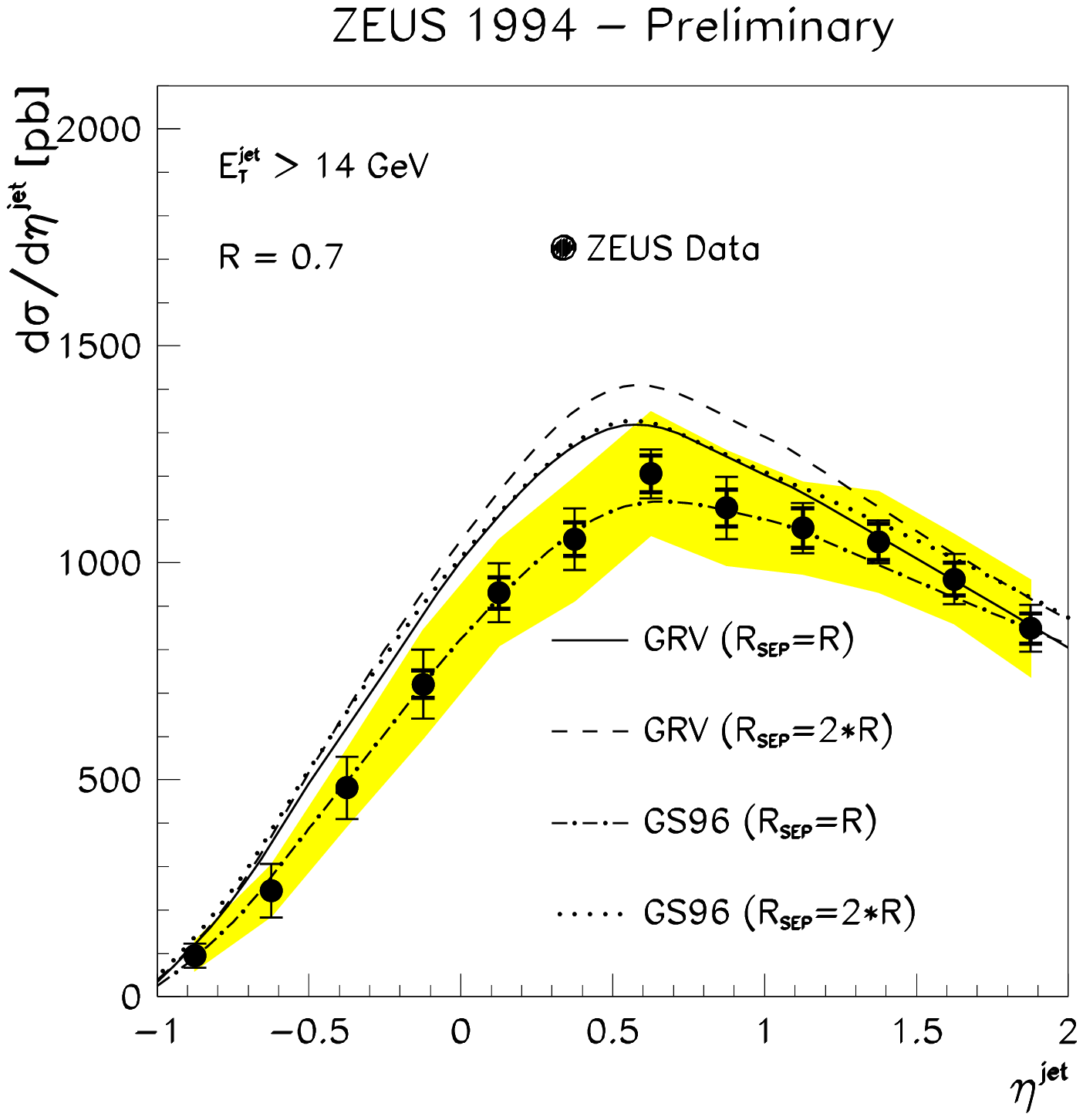,width=7.5cm}}}
\vspace{-1.5cm}
\caption{Inclusive jet differential cross sections.
\label{fig:two}}
\vspace{-0.5cm}
\end{figure}

Figure \ref{fig:two} shows the measurements of the inclusive jet differential
cross sections for jets searched with an iterative cone algorithm
\cite{p:cone} as a function of the jet pseudorapidity ($\etajet$) for jets
with transverse energy satisfying $\etjet>14$ GeV. The measurements were
performed in the kinematic region given by $0.2<y<0.85$ and $\q2\leq 4$ \g2,
and for two cone radii: $\R71$. The behaviour of the cross section as a
function of $\etajet$ in the region $\etajet>1$ is very different for the two
cone radii: it is flat for $\rr1$ whereas it decreases as $\etajet$ increases
for $\r7$. It is noted that a given $\etjet$ threshold for jets defined with
$\r7$ corresponds to a higher $\etjet$ threshold for jets with $\rr1$ and this
fact may account for part of the observed differences.

Next-to-leading order (NLO) QCD calculations \cite{p:kramer3} are compared to
the measurements in figure \ref{fig:two} using two different parametrisations
of the photon structure function: GRV-HO \cite{p:grvho} and GS96 \cite{p:gs},
and two values of \footnote{The parameter \cite{p:sdellis} $\rs$ is
introduced into the NLO calculations in order to simulate the experimental jet
algorithm by adjusting the minimum distance in $\etaphi$ at which two partons
are no longer merged into a single jet.} $\rs$. The CTEQ4M \cite{p:cteq4}
proton parton densities have been used in all cases. For forward jets with
$\rr1$ an excess of the measurements with respect to the calculations is
observed. This discrepancy is attributed to a possible contribution from
non-perturbative effects (e.g., the so-called ``underlying event''), which are
not included in the theoretical calculations. This contribution gets largely
reduced by decreasing the size of the cone since the transverse energy density
inside the cone of the jet due to the underlying event is expected to be
roughly proportional to the area covered by the cone. A very good agreement
between data and NLO calculations is observed for measurements performed using
a cone radius of $\r7$ for the entire $\etajet$ range measured. The
predictions using GRV-HO and GS96 show differences which are of the order of
the largest systematic uncertainty of the measurements. Thus, these
measurements exhibit a sensitivity to the parton densities in the photon and
can be used in quantitative studies.

To study the internal structure of the jets, the jet shape $\psi(r)$ has been
used. $\psi(r)$ is defined as the average fraction of the jet's transverse
energy that lies inside an inner cone of radius $r$, concentric with the jet
defining cone~\cite{p:sdellis}:

\begin{equation}
\psi(r) = \frac{1}{N_{jets}} \sum_{jets} \frac{E_T(r)}{E_T(r=R)}
\label{eq:three}
\end{equation}
where $E_T(r)$ is the transverse energy within the inner cone and $N_{jets}$ is
the total number of jets in the sample. By definition, $\psi(r=R)=1$. The jet
shape is affected by fragmentation and gluon radiation. However, at
sufficiently high $\etjet$ the most important contribution is predicted to come
from gluon emission off the primary parton.

\begin{figure}
\vspace{-1cm}
\centerline{\mbox{
\epsfig{figure=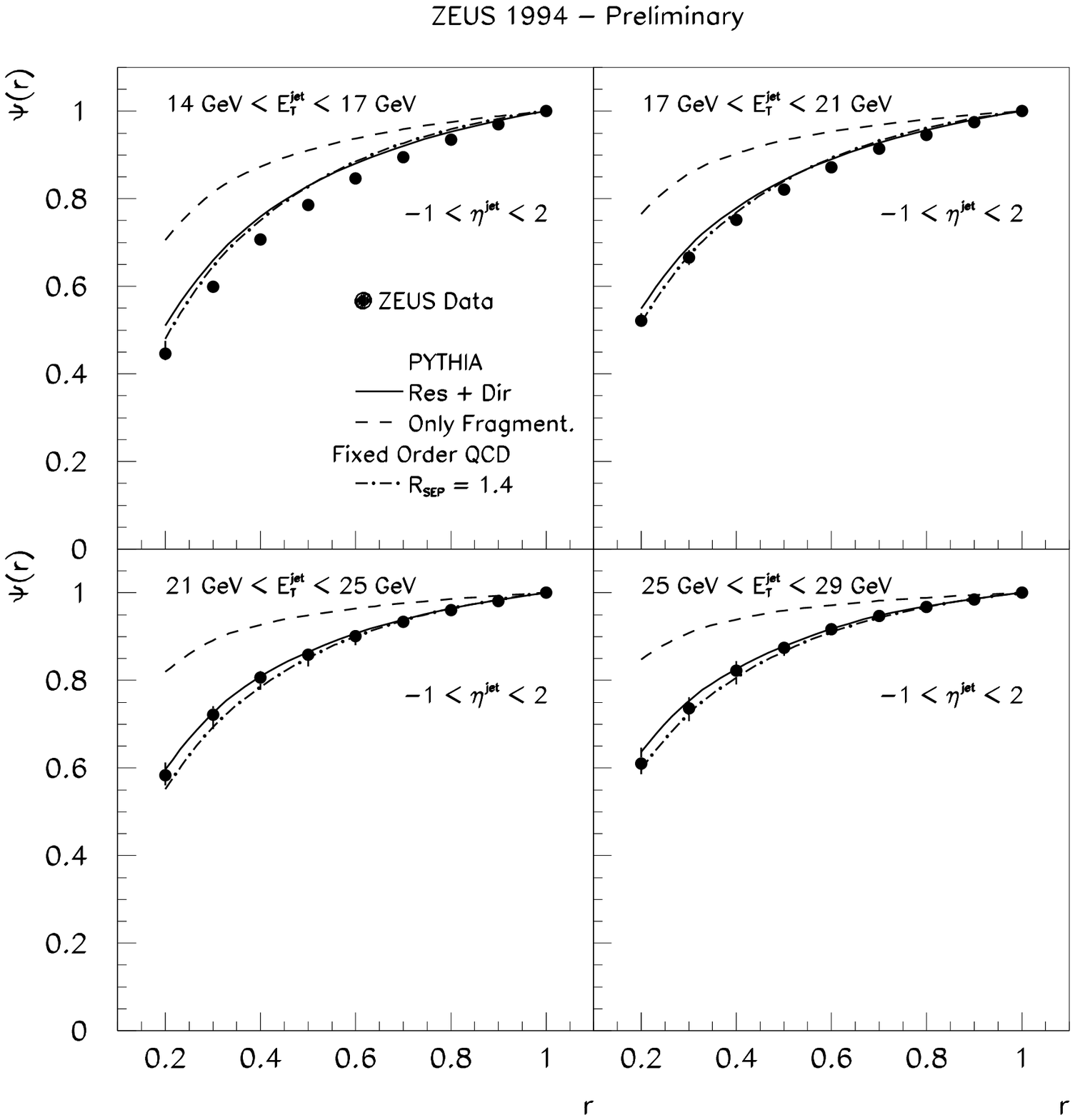,width=9cm}}}
\vspace{-0.5cm}
\caption{Jet shapes: $\etjet$ dependence.
\label{fig:three}}
\vspace{-0.5cm}
\end{figure}

Figure \ref{fig:three} shows the measured jet shapes $\psi(r)$ as a function
of the inner cone radius $r$ using a cone algorithm with radius $\rr1$ for
jets with $\etar$ and in four regions of $\etjet$. As a jet becomes narrower,
the value of $\psi(r)$ increases for a fixed value of $r$. It is observed (see
figure \ref{fig:three}) that the jets become narrower as $\etjet$ increases.
For comparison, the predictions from leading-logarithm parton-shower Monte
Carlo calculations as implemented in the PYTHIA program \cite{p:pythia} for
resolved plus direct processes are shown in the same figure. For $\etjet>17$
GeV the predictions reproduce reasonably well the data. In the lowest-$\etjet$
region small differences between data and the predictions are observed. PYTHIA
including resolved plus direct processes but without initial and final state
parton radiation predicts jet shapes which are too narrow in each region of
$\etjet$. These comparisons show that parton radiation is the dominant
mechanism responsible for the jet shape in the whole range of $\etjet$ studied.

\begin{figure}
\vspace{-1cm}
\centerline{\mbox{
\epsfig{figure=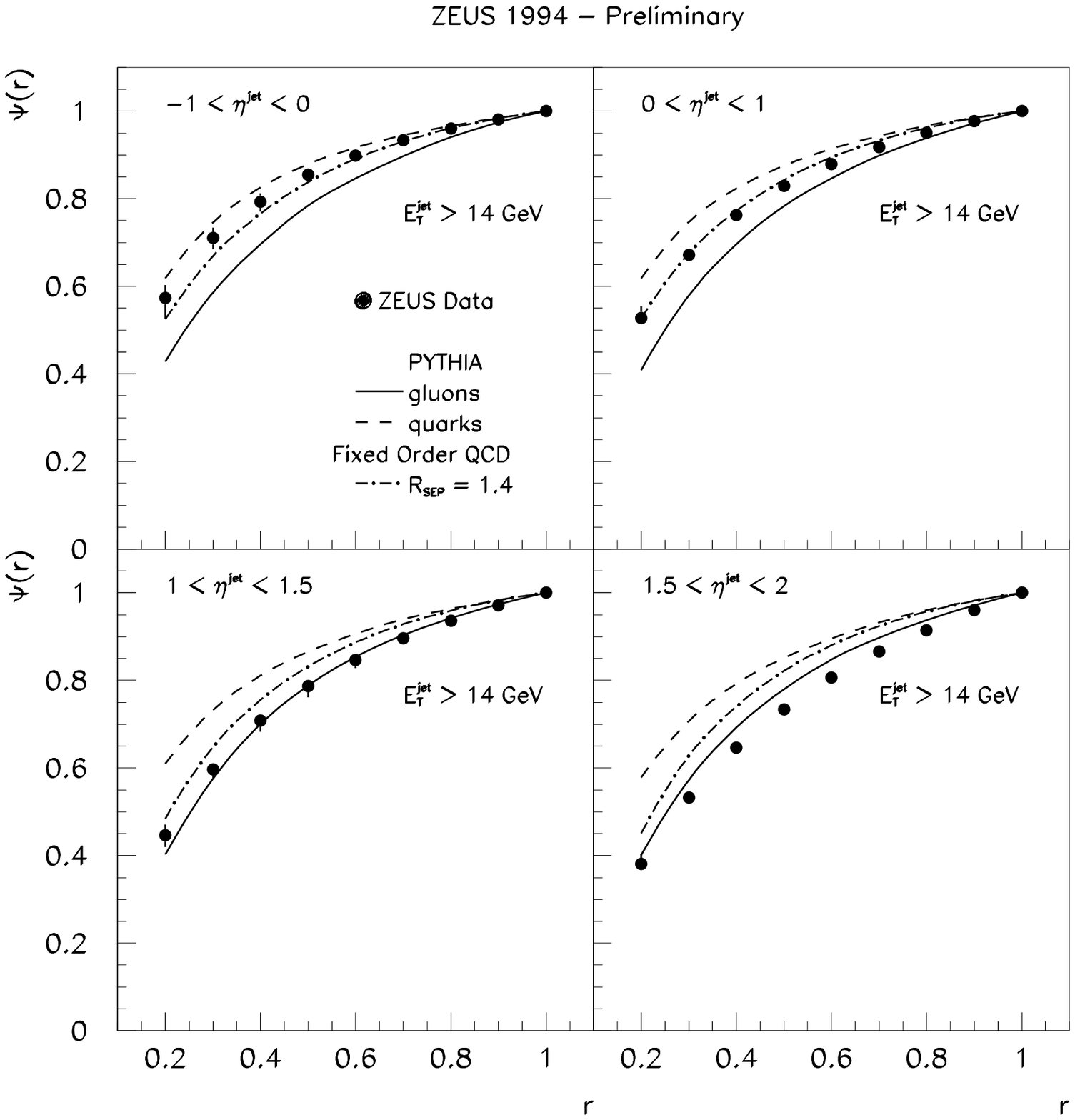,width=9cm}}}
\vspace{-0.5cm}
\caption{Jet shapes: $\etajet$ dependence.
\label{fig:four}}
\vspace{-0.5cm}
\end{figure}

The $\etajet$ dependence of the jet shape is presented in figure
\ref{fig:four}. It is observed that the jets become broader as $\etajet$
increases. Perturbative QCD predicts that gluon jets are broader than quark
jets as a consequence of the fact that the gluon-gluon is larger than the
quark-gluon coupling strength. The predictions of PYTHIA for quark and gluon
jets are shown in figure \ref{fig:four}. The data go from being dominated
by quark jets in the final state ($\etajet<0$) to being dominated by gluon
jets ($\etajet>1$). Therefore, the broadening of the measured jet shapes as
$\etajet$ increases is consistent with an increase of the fraction of gluon
jets.

Lowest non-trivial-order QCD calculations \cite{p:kramer} of the jet shapes
are compared to the measurements in figures \ref{fig:three} and
\ref{fig:four}. The fixed-order QCD calculations with a common value of
$\rs=1.4$ reproduce reasonably well the measured jet shapes in the region
$\etjet>17$ GeV and in the region $-1<\etajet<1$.

The results on photoproduction of jets presented here constitute a
step forward towards testing QCD and the extraction of the photon parton
densities.

\vspace{-0.2cm}
\section*{Acknowledgments}
\vspace{-0.3cm}
I would like to thank the organizers of the conference for an enriching
atmosphere and their warm hospitality.

\end{document}